\documentclass[12pt]{article}
\usepackage{amsfonts,amsmath,epsfig,graphicx,amssymb}
\usepackage{amsmath,epsfig,graphicx,amssymb}
\usepackage[margin=1in]{geometry}
\usepackage{amsmath}
\usepackage{amsfonts}
\usepackage{amssymb}
\usepackage{makeidx}
\usepackage{graphicx}
\newcommand{\beq}{\begin{equation}}
\newcommand{\eeq}{\end{equation}} 
\newcommand{\ben}{\begin{eqnarray}}
\newcommand{\een}{\end{eqnarray}}
\date{}
\begin{document}
\title{The Story of Bose, Photon Spin and Indistinguishability}
\author{Partha Ghose\footnote{partha.ghose@gmail.com} \\
Tagore Centre for Natural Sciences and Philosophy,\\ Rabindra Tirtha, New Town, Kolkata 700156, India}
\maketitle
\begin{abstract}
As we approach the centenary of the discovery of quantum statistics in 1924, it is important to revisit Bose's original derivation of Planck's law usually ignored in most standard presentations of Bose-Einstein statistics. It introduced not only the novel concept of the {\em indistinguishability} of photons but also of their intrinsic {\em spin}, a fact unknown to most physicists.    
\end{abstract}
By 1923 Satyendra Nath Bose (born 1st January 1894) had already made his mark as a prodigious talent in the intellectual circles of Calcutta (now Kolkata), the capital of British India, but was still a total stranger in the world of physics at large. In the early 1920s he was teaching quantum theory to students at the newly established Dacca University (now called Dhaka University) in erstwhile East Bengal which eventually became Bangladesh in 1972. He was struggling with the derivation of Planck's law -- he found all previous derivations, including those by Einstein, to be logically unsatisfactory because of the simultaneous use of both classical and quantum ideas. He was looking for a derivation that was entirely quantum theoretical. 

On June 4, 1924 he sent a short paper ‘Planck’s Law and the Light-Quantum Hypothesis’ to Albert Einstein with the humble request, ‘You will see that I have tried to deduce the coefficient $8\pi\nu^2/c^3$ in Planck’s law independent of the classical electrodynamics, only assuming that the ultimate elementary regions in the Phase space have the content $h^3$. I do not know sufficient German to translate the paper. If you think the paper worth publication, I shall be grateful if you arrange its publication in Zeitschrift f\"{u}r Physik’. 

In a post card dated 2nd July, 1924 Einstein wrote to Bose, ‘Dear Colleaugue, I have translated your work ... It signifies an important step forward and I liked it very much ...... You are the first to derive the factor quantum theoretically, even though because of the polarization factor 2 not wholly rigorously. It is a beautiful step forward’. In a footnote to the translated paper, Einstein wrote: ‘In my opinion Bose’s derivation signifies an important advance. The method used
here gives the quantum theory of an ideal gas as I will work out elsewhere. -A Einstein’ [1]. Ten days later on 12th July he wrote to Ehrenfest, ‘The Indian Bose has given a beautiful derivation of Planck’s law including the constant $8\pi\nu^2/c^3$’.

With such a strong endorsement from Einstein, Bose instantly emerged as a prominent figure in the world of quantum physics. His biography and the impact of his  paper on physics as a whole can be gleaned from [2, 3, 4]. It would be instructive to revisit Bose’s original paper, particularly since standard presentations of Bose-Einstein statistics gloss over two of its revolutionary insights, both of which are contained in Bose’s ingenious derivation of
the phase space factor $8\pi\nu^2/c^3$, namely the idea of {\em indistinguishability} of photons which is usually presented as an {\em ad hoc} assumption and the idea of {\em an additional internal degree of freedom of light-quanta} which is totally ignored.
  
Let's begin with the Planck formula for the spectral density of electromagnetic radiation emitted by a black body in thermal equilibrium at a given temperature $T$,

\beq 
\rho(\nu, T) =\left(\frac{8\pi\nu^2 d\nu}{c^3}\right)\left(\frac{h\nu}{e^{h\nu / kT} - 1}\right).\label{Pl}
\eeq 
Planck tried very hard to find a theoretical basis for it, and finally `as an act of desperation.... to obtain a positive result, under any circumstances and at whatever cost' introduced the idea of quanta of energy for the material in the oven walls, {\em not in the radiation itself} which he took to be classical and described by Maxwell's theory. In mid-December 1900 he presented a statistical derivation involving a distribution of the hypothetisized energy quanta among the wall `resonators' that had no more justification than that it gave the right result.

With a hunch that the law implied a non-classical corpuscular nature of radiation itself, Einstein used some results of Wien on the entropy of radiation to calculate the volume dependence of the entropy of thermal radiation, and from that he drew the following revolutionary conclusion:
\begin{quote}
 ... We (further) conclude that monochromatic radiation of low density (within the range of validity of Wien's radiation formula) behaves thermodynamically as if it consisted of {\em mutually independent energy quanta} of magnitude $R\beta\nu/ N$. (italics added) 
\end{quote}
The factor $R\beta\nu/N = h\nu$.  The idea of light-quanta was thus born, with all its problems. First, there was a fundamental conflict between the statistical independence of Einstein's light-quanta and the Planck law which implies correlated quanta \cite{nat, ehr}. Second, there was no {\em logically satisfactory} derivation of the Planck law as a whole despite several attempts by the leading physicists of the time suh as Planck himself \cite{pl}, Peter Debye \cite{deb}, Einstein \cite{ein}, Pauli \cite{pau} and Einstein and Ehrenfest \cite{ehren}. In all of them the first factor $(8\pi\nu^2 d\nu/c^3)$ in the Planck law (\ref{Pl}) was taken from classical electrodynamics to be the number density of the modes of vibration of radiation. The second factor was deduced by postulating various {\em ad hoc} rules. Third, in all these attempts (except that of Debye) Planck's hypothesis, namely that quantization was restricted to the exchange of energy between radiation and matter, was used. But by late 1923/early 24 the Compton Effect had clearly shown that radiation itself consists of energy quanta. 

Bose solved all these problems in one stroke by deriving the full Planck law, including the first factor, from quantum theory. As we have seen, in his letter to Einstein, he wrote, `You will see that I have tried to deduce the coefficient $8\pi\nu^2/c^3$ independent of the classical electrodynamics.' This is never mentioned in text books and other expositions, but is of crucial importance. He did that by extending Planck's method of quantization of material resonators to radiation itself. Consider the phase space of a single photon,
\beq
\int dV . dp_x dp_y dp_z = V 4\pi p^2 dp = V\frac{4\pi h^3\nu^2 d\nu} {c^3}
\eeq
on using the Einstein relation $p = h\nu/c$. Since the irreducible elementary regions of phase space are of size $h^3$, dividing by $Vh^3$ one obtains the number density of phase space cells,
\beq
\frac{4\pi\nu^2 d\nu} {c^3}
\eeq
It is such a simple derivation, except that it is wrong by a factor of 2. The paper is slightly hesitant at this point and says, `It seems, however, appropriate to multiply this number once again by 2 in order to take into account the fact of polarization.' Therein hangs a tale. But before going into that, suffice it to say here that the first factor now emerges with a totally new meaning: it is the number of possible states of the photon, i.e., the number of possible arrangements of a photon in the given volume. This number is fixed, and hence permutations of the (identical) photons within a cell cannot produce new cells, and this immediately implies that the photons are {\em indistinguishable particles}. The cells are therefore labelled only by the number of photons in them. Let $p^s_i$ be the number of cells with $i$ number of photons, each with frequency in the range $\nu^s$ and $\nu^s + d\nu^s$. Then the probability of a state with type $s$ quanta is 
\beq
W^s    = \frac{A^s!}{ p^s_0 !\, p^s_1 !\, p^s_2 !\, \cdots} \label{1}
\eeq
with
\beq
A^s    = \frac{8\pi \nu^{s2} d\nu^s}{c^3}
\eeq
and 
\beq
N^s =  0. p^s_0 + 1. p^s_1 + 2. p^s_2 + \cdots
\eeq
The macroscopically defined probability of a state defined by all types of quanta is then
\begin{equation}
 W = \Pi_s W^s = \Pi_s \frac{A^s!}{ p^s_0 !\, p^s_1 !\, p^s_2 !\, \cdots}.\label{bose}
\end{equation} 
Maximizing the entropy $S = k {\rm log} W$ subject to the constraints that the total energy is the sum of the energies of all the quanta, $E = \Sigma_ s N^s h\nu^s$, and the total number of quanta is the sum of the numbers of quanta in all the states, $N^s =  \Sigma_r r p^s_r$, leads then, in the limit of large numbers $p^s_r$, to the statistical Planck factor $(1 - e^{- h\nu/ kT})^{- 1}$. 
QED. No hypothetical energy quanta, no wall resonators, just physical indistinguishable photons! One of the foundations of quantum mechanics was laid, and it followed rapidly.

Now to the story behind the factor of 2 which signals {\em an additional degree of freedom of the photon}. I cannot recollect the exact date, but most probably sometime in the late 70's I had gone to meet Bose in his residence. We were alone in his living room. He told me the following story in hush hush confidence and made me promise never to divulge it publicly. Why then am I breaking my promise? It'll become clear shortly. `You know', he said,  `my deduction of the Planck law had a factor of 2 missing. So I proposed that it came from the fact that the photon had a spin, and that it can spin either parallel or antiparallel to its direction of motion. That would give the additional factor of 2. But the old man (meaning Einstein) crossed it out (`budho k\'{e}t\'{e} dil\'{e}' in Bengali, his mother tongue) and said it was not necessary to talk about spin, the factor of 2 comes from the two states of polarization of light.' Then he said rhetorically with a mischievous distant look, `I can understand a spinning particle, but what can the polarization of a particle mean?' I was stunned! 

Much later (again I cannot recollect the exact date, but certainly sometime during 1993 when we were busy preparing for his birth centenary celebrations in January 1994) I accidentally chanced upon a 1931 paper of Raman and Bhagavantam with the title ``Experimental proof of the spin of the photon'' \cite{ram}. I was intrigued and started reading it. I was amazed. Let me quote:
\begin{quote}
In his well-known derivation of the Planck radiation formula from quantum statistics, Prof. S. N. Bose obtained an expression for the number of cells in phase space occupied by the radiation, and found himself obliged to multiply it by a numerical factor 2 in order to derive from it the correct number of possible arrangements of the quantum in unit volume. The paper as published did not contain a detailed discussion of the necessity for the introduction of this factor, but we understand from a personal communication by Prof. Bose that he envisaged the possibility of the quantum possessing besides energy $h\nu$ and momentum $h\nu/c$ also an intrinsic angular momentum $\pm h/2\pi$ around an axis parallel to he direction of its motion. The weight factor 2 thus arises from the possibility of the spin of the quantum being right-handed or left-handed, corresponding to the two alternative signs of the angular momentum. There is a fundamental difference between this idea, and the well-known result of classical electrodynamics to which attention was drawn by Poynting and more fully developed by Abraham that a beam of light may in certain circumstances possess angular momentum. Thus, according to the classical field theory, the angular momentum associated with a quantum of energy is not uniquely defined, while according to the view we are concerned with in the present paper, the photon has always an angular momentum having a definite numerical value of a Bohr unit with one or the other of the two possible alternative signs.
\end{quote}

The Raman-Bhagavantam experiment conclusively showed that Bose's view was the correct one. It was {\em the first ever experimental determination of photon spin}, but how many scientists and historians of science know this? 

It should be clear now why I feel no compunction in breaking my promise--the story has been in the public domain since 1931, though hardly noticed by anyone! Did Bose forget about this paper? Unlikely. By the time I unearthed it, he was no more, so I couldn't check back with him. I guess it must have been due to his deep respect for and gratitude to Einstein. After all he owed his entire reputation to Einstein. There were strong initial objections to Bose's method. For example, Ehrenfest had objected to the new statistics on the ground that the `mysterious influences' among the molecules was inconsistent with their statistical independence. Einstein's response was characteristic--Bose's counting method was the only one consistent with Nernst's heat theorem for ideal gases. He wrote \cite{eins1}, `one recognizes at once that the entropy must vanish at the absolute zero of temperature. The reason is that then all molecules are in the first cell; and for this state there exists only one distribution of molecules according to our counting method. Hence our assertion is immediately proved to be correct.' And again \cite{eins2}, `This result represents in itself a support of the view concerning the deep natural relation between radiation and gas, since the same statistical treatment which leads to Planck's formula establishes -- when applied to ideal gases-- the agreement with Nernst's theorem.' Thus was born the idea of Bose-Einstein Condensates. Einstein therefore proposed to proceed with Bose's method, brushing aside all objections and leaving the problem of `mysterious influences' among the molecules to be sorted out in the future. 

To conclude, the impact of Bose's short paper turned out to be far-reaching. In particular it gave birth to the field of quantum statistics-- a seminal and essential part of quantum theory. In his book `Subtle is the Lord' Pais states, `The paper by Bose is the fourth and last of the revolutionary papers of the old quantum theory (the other three being by, respectively, Planck, Einstein and Bohr). As noted above, Bose's idea of ``indistinguishability'', followed by Einstein's essential contributions, gave rise to the idea of Bose-Einstein Condensates, which now has numerous applications. Last but not least, as we now know, nature has only two types of particles - bosons and fermions - depending upon whether they possess integer or half-integer spin respectively, a fact that follows as a general consequence of relativistic quantum field theory, and is of course amply verified empirically'.

A short paper by an unknown physicist from an unknown University in a far off British Colony contained two seminal ideas -- the idea of {\em indistinguishability} of photons and the idea of {\em an additional degree of freedom of photons} which later came to be called `spin'.

This is certainly a story worth telling.

I am grateful to Jogesh Pati for encouraging me to write this article and suggesting several improvements of the original draft.

\end{document}